\documentclass[prb,twocolumn,amsmath,amssymb,floatfix,superscriptaddress,aps]{revtex4-1}
\usepackage[utf8]{inputenc}
\usepackage{graphicx}
\usepackage{dcolumn}
\usepackage{bm}
\usepackage{color}
\usepackage{soul}
\usepackage{cancel}

\usepackage{lipsum}
\usepackage{comment}
\usepackage{braket}
\usepackage{mathalfa}
\usepackage[caption = false]{subfig}
\newcommand{\be}{\begin{equation}}
\newcommand{\ee}{\end{equation}}
\newcommand{\bea}{\begin{eqnarray}}
\newcommand{\eea}{\end{eqnarray}}
\newcommand{\ba}{\begin{align}}
\newcommand{\ea}{\end{align}}

\usepackage[capitalise]{cleveref}

\begin{document}

\title{Thermoelectric efficiency in multiterminal quantum thermal machines
  from steady-state density functional theory}

\author{N. Sobrino}
\email{nahualcsc@dipc.org}
\affiliation{Nano-Bio Spectroscopy Group and European Theoretical Spectroscopy Facility (ETSF), Departamento de Pol\'imeros y Materiales Avanzados: F\'isica, Qu\'imica y Tecnolog\'ia, Universidad del Pa\'is Vasco UPV/EHU, Avenida de Tolosa 72, E-20018 San Sebasti\'an, Spain}

\author{R. D'Agosta}
\email{roberto.dagosta@ehu.es}
\affiliation{Nano-Bio Spectroscopy Group and European Theoretical Spectroscopy Facility (ETSF), Departamento de Pol\'imeros y Materiales Avanzados: F\'isica, Qu\'imica y Tecnolog\'ia, Universidad del Pa\'is Vasco UPV/EHU, Avenida de Tolosa 72, E-20018 San Sebasti\'an, Spain}
\affiliation{IKERBASQUE, Basque Foundation for Science, Plaza de Euskadi 5, E-48009 Bilbao, Spain}

\author{S. Kurth}
\email{stefan.kurth@ehu.es}
\affiliation{Nano-Bio Spectroscopy Group and European Theoretical Spectroscopy Facility (ETSF), Departamento de Pol\'imeros y Materiales Avanzados: F\'isica, Qu\'imica y Tecnolog\'ia, Universidad del Pa\'is Vasco UPV/EHU, Avenida de Tolosa 72, E-20018 San Sebasti\'an, Spain}
\affiliation{IKERBASQUE, Basque Foundation for Science, Plaza de Euskadi 5, E-48009 Bilbao, Spain}
\affiliation{Donostia International Physics Center, Paseo Manuel de Lardizabal 4, E-20018 San Sebasti\'an, Spain}

\date{\today}
\begin{abstract}
  The multi-terminal generalization of the steady-state density functional
  theory for the description of electronic and thermal transport (iq-DFT) is
  presented. The linear response regime of the framework is developed  leading
  to exact expressions for 
  the many-body transport coefficients and thermoelectric efficiency purely
  in terms of quantities accessible to the framework. The theory is applied to
  a multi-terminal interacting quantum dot in the Coulomb blockade regime for
  which accurate parametrizations of the exchange-correlation kernel matrix are
  given. The thermoelectric efficiency and output power of the multi-terminal
  system are studied. Surprisingly, the strong-interaction limit of these
  quantities can be understood in terms of the non-interacting one.
\end{abstract}

\maketitle
 
\section{Introduction}
In recent decades, electronic transport through nanoscale devices, even
down to the size of single molecules, has attracted increasing scientific
and technological interest.
\cite{cuniberti2006introducing,cuevas2010molecular} The main motivation for
this interest is the reduction of the dimensions of active electronic
devices, e.g., control of electronic currents at ever smaller scales.
However, at these small scales, heat management becomes crucial for
reliable device operation. Also, one may aim to harness
thermal energy by conversion to electrical currents in thermoelectric
devices. \cite{goldsmid2010introduction,sanchez2011optimal,benenti2017fundamental,erdman2017thermoelectric,sothmann2014thermoelectric,whitney2014most,EsfarjaniZebarjadiKawazoe:06}
Therefore it is important to deal with both electrical and thermal transport
on equal footing. Although two-terminal setups have been the focus of most
investigations, the exploration of thermoelectric transport in multi-terminal devices \cite{jacquet2009thermoelectric,entin2010three,sanchez2011thermoelectric,jiang2012thermoelectric,SahaLuBernholcMeunier:09,sothmann2012magnon,brandner2013strong,balachandran2013efficiency,mazza2014thermoelectric} has started more
recently due to the potential added benefits of these more intricate designs,  to, e.g., 
separate heat and electrical transport.

For non-interacting electrons, an adequate framework to describe both electronic
and heat transport in the steady state is the Landauer-B\"uttiker (LB)
formalism \cite{buttiker1985generalized,buttiker1986four} which treats
transport essentially as a scattering problem. On the other hand, to
describe the currents through an interacting region attached to
non-interacting leads, we have the Meir-Wingreen formula
\cite{meir1992landauer}, which expresses the currents in terms of the
many-body spectral function. 

For an ab-initio modeling of materials, density functional theory
(DFT) \cite{kohn1965self} has become an indispensable tool, mainly
due to its reasonable balance between accuracy and numerical efficiency.
While DFT was originally formulated for (thermal) equilibrium, a combination
of DFT with the LB formalism has widely been used to
model both electronic and heat transport through, e.g., single molecules
\cite{Lang:95,DiVentraPantelidesLang:00,TaylorGuoWang:01,BrandbygeMozosOrdejonTaylorStokbro:02,NikolicSahaMarkussenThygesen:12,ThossEvers:18}.
However, one has to keep in mind that standard DFT is not designed to
describe out-of-equilibrium physics such as electronic or thermal transport.
While in special circumstances this may be enough to capture, e.g., linear
transport coefficients \cite{stefanucci2011towards,bergfield2012bethe,troster2012transport,yang2016density},
in general extensions of the theory are required. One such possible extension
for the description of transport is steady-state DFT (or i-DFT) 
\cite{stefanucci2015steady} which adds to the basic quantity of standard DFT
and LB-DFT, the density, another fundamental variable, the steady-state
electronic current. This framework has been used to describe transport
through model systems, including strongly correlated ones
\cite{kurth2016nonequilibrium,kurth2017transport,jacob2018many,jacob2020mott,sobrino2020exchange}
and has also been formulated to deal with multi-terminal systems.
\cite{kurth2019nonequilibrium}

In more recent work \cite{sobrino2021thermoelectric}, an extension of
steady-state DFT, dubbed iq-DFT, has been suggested which besides the
electronic (particle) current also allows for the description of
(electronic) heat or energy currents. In the present work, we extend
iq-DFT to systems connected to an arbitrary number of leads
(Sec.~\ref{section_formalism}) with explicit development of the
linear-response regime. In Sec.~\ref{siam} A, the formalism is applied to the
single-impurity Anderson model (SIAM) in the Coulomb blockade regime for
which we present the exchange-correlation (xc) kernel of linear response.
This allows to study all the linear transport coefficients (electrical
and thermal conductances, Seebeck coefficients, etc.) solely in terms of
iq-DFT quantities. For the numerical results (Sec.~\ref{results_num}) we
focus on the multi-terminal efficiency of the SIAM viewed as a thermal
machine and we show explicitly that in the strong-interaction limit this
quantity strictly reduces to its non-interacting counterpart. Finally, we
present our conclusions in Sec.~\ref{conclus}.

\section{Multi-terminal iq-DFT}
\label{section_formalism}
\begin{figure}
	\centering
	\includegraphics[width=0.9\linewidth]{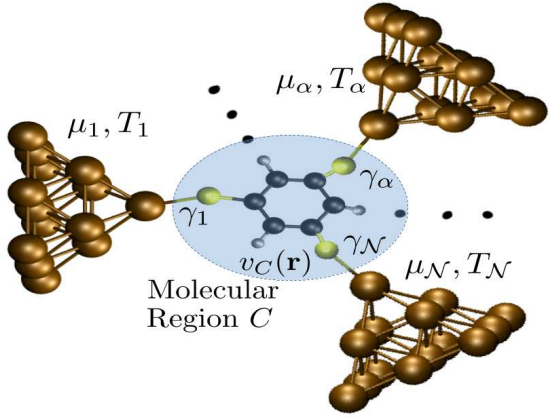}
	\caption{ Schematic drawing of a multi-terminal nanoscale
		junction. A molecular region  $C$ at gate potential $v_{C}(\textbf{r})$ is coupled to $\mathcal{N}$
		leads at chemical potentials $\mu_{\alpha}=\mu+V_{\alpha}$  and temperatures $T_{\alpha}=T(1+\Psi_{\alpha})$.}
	\label{fig_1}
\end{figure}

We consider a general electronic transport setup as depicted in \cref{fig_1},
where $\mathcal{N}$ (semi-infinite) electrodes are coupled to a central
(molecular) region ($C$) subject to an electrostatic potential
$v_{C}(\mathbf{r})$ which vanishes deep inside the electrodes (leads). While the
electrodes are assumed to be at {\em local} equilibrium characterized by 
temperatures $T_{\alpha}$ and chemical potentials $\mu_{\alpha}$
($\alpha=1,\dots,\mathcal{N}$), the total system is considered to be in a
non-equilibrium steady state. For convenience we also define both an
{\em equilibrium} temperature and chemical potential as
$T=\frac{1}{\cal{N}}\sum_{\alpha} T_{\alpha}$ and
$\mu=\frac{1}{\cal{N}}\sum_{\alpha} \mu_{\alpha}$, respectively, such that the
the lead temperatures $T_{\alpha}=T(1+\Psi_{\alpha})$ can be expressed in terms
of thermal gradients $\Psi_{\alpha}$ while the chemical potentials
$\mu_{\alpha}=\mu+V_{\alpha}$ are written in terms of DC biases $V_{\alpha}$. From
these definitions of $\mu$ and $T$ it follows that $\sum_{\alpha}V_{\alpha}=0$ and
$\sum_{\alpha}\Psi_{\alpha}=0$, respectively.

The non-equilibrium steady state of the system is characterized by the
electronic density $n(\mathbf{r})$ in region $C$, as well as two sets of
steady currents: (a) the electronic and energy currents ($I_{\alpha}$ and
$W_{\alpha}$, respectively) flowing from lead $\alpha$ to region $C$, or
(b) the corresponding electronic and heat currents ($I_{\alpha}$ and
$Q_{\alpha}$, respectively). These currents are related through
\begin{align}
	W_{\alpha}= Q_{\alpha}+\mu_{\alpha} I_{\alpha}
	\label{eq_mult_W_to_Q},
\end{align}
and the muti-terminal iq-DFT approach described below can equivalently be
formulated in terms of both fundamental current variables (a) or (b).

In the following, we adopt the sign convention that currents flowing into the
central region $C$ are positive. Due to charge and energy conservation, we have
$\sum_{\alpha} I_{\alpha}=0$, $\sum_{\alpha} W_{\alpha}=0$, while from
Eq.~(\ref{eq_mult_W_to_Q}) we obtain for the heat currents
$\sum_{\alpha} Q_{\alpha}=-\sum_\alpha I_{\alpha}V_{\alpha}$.
 Furthermore, atomic units are used throughout. 
Energies are given in units of temperature unless otherwise noted.

The extension of the iq-DFT formalism \cite{sobrino2021thermoelectric} to
multi-terminal setups can be formally established through the following
theorem. Here, without loss of generality, we assume that the gradients and currents associated with the $\mathcal{N}$th lead are expressed in terms of the gradients and currents of the other $\mathcal{N}-1$ leads.  

{\em Theorem}: 
There exists a one-to-one correspondence between the set of ``densities''
($n(\mathbf{r}), I_{1},Q_{1},\dots,I_{\mathcal{N}-1}, Q_{\mathcal{N}-1}$)  and the set of ``potentials'' ($v(\mathbf{r}), V_{1}/T,\Psi_{1}/T,\dots,V_{\mathcal{N}-1}/T,\Psi_{\mathcal{N}-1}/T$), for any finite temperature $T$ and fixed electrostatic potential in the leads,
in a finite region around
$V_{\alpha}=0$ and $\Psi_{\alpha}=0$ for all $\alpha=1,\dots,\mathcal{N}-1$. The proof of the theorem is analogous to the one presented in
Ref.~\onlinecite{sobrino2021phdthesis}.

According to the theorem and under the usual assumption of
non-interacting representability, there exists a unique set of Kohn-Sham (KS)
potentials $(v_{s}(\mathbf{r}), V_{s,1}/T,\Psi_{s,1}/T,\ldots, V_{s,\mathcal{N}-1}/T, 
\Psi_{s,\mathcal{N}-1}/T)$ which in a noninteracting system reproduces the density
$n(\mathbf{r})$ and currents ($I_{1},Q_{1},\dots,I_{\mathcal{N}-1},Q_{\mathcal{N}-1})$
of the interacting system.
Following the standard KS procedure, the xc potentials are defined as 
\begin{subequations}
	\begin{align}
	  v_{\rm Hxc}[n,\mathcal{I},\mathcal{Q}](\mathbf{r}) &
          =v_{s}[n,\mathcal{I},\mathcal{Q}](\mathbf{r}) -
          v[n,\mathcal{I},\mathcal{Q}](\mathbf{r}) ,\\
	  V_{{\rm xc},\alpha}[n,I,Q] &=
          V_{s,\alpha}[n,\mathcal{I},\mathcal{Q}] -V_\alpha[n,\mathcal{I},\mathcal{Q}], \\
	  \Psi_{{\rm xc},\alpha}[n,\mathcal{I},\mathcal{Q}]&=
          \Psi_{s,\alpha}[n,\mathcal{I},\mathcal{Q}] -\Psi_\alpha[n,\mathcal{I},\mathcal{Q}],
	\end{align}
	\label{eq_XC_potentials}
\end{subequations}
for $\alpha=1,\dots, \mathcal{N}-1$, where $\mathcal{I} = (I_1,\ldots,I_{\mathcal{N}-1})$ and $\mathcal{Q} = (Q_1,\ldots,Q_{\mathcal{N}-1})$.
The self-consistent coupled KS equations for the densities read
($\int\equiv\int_{-\infty}^{\infty} \frac{d\omega}{2\pi}$ in the following)
\begin{subequations}
	\begin{align}
	  &n(\mathbf{r})=2\sum_{\alpha=1}^{\cal{N}}\int f(\omega_{s,\alpha})
          A_{s,\alpha}(\mathbf{r},\omega)\label{eq_KS_densities_n},\\
	  &I_{\alpha}=2\sum_{\alpha'=1}^{\cal{N}}\int \left[ f(\omega_{s,\alpha})-
            f(\omega_{s,\alpha'})\right] \mathcal{T}_{s,\alpha\alpha'}(\omega)
          \label{eq_KS_densities_I},\\
	  &Q_{\alpha}=2\sum_{\alpha'=1}^{\cal{N}}\int \left[ f(\omega_{s,\alpha})-
            f(\omega_{s,\alpha'}) \right](\omega-\mu_{s,\alpha})
          \mathcal{T}_{s,\alpha\alpha'}(\omega)\, ,\label{eq_KS_densities_Q}
	\end{align}
	\label{eq_KS_densities}
\end{subequations}
where $\omega_{s,\alpha}=\frac{w-\mu_{s,\alpha}}{1 + \Psi_{s,\alpha}}$ with  
$f(z) = [1 + \exp(z/T)]^{-1}$ being the Fermi function and  
$\mu_{s,\alpha}=\mu+V_{s,\alpha}$ with the KS bias for lead $\alpha$
\begin{equation}
  V_{s,\alpha} = V_{\alpha}+V_{xc,\alpha}[n,\mathcal{I},\mathcal{Q}] \;.
\end{equation}
In \cref{eq_KS_densities}, the (partial) KS spectral function is defined as 
$A_{s,\alpha}(\mathbf{r},\omega)=\bra{\mathbf{r}}\mathcal{G}(\omega)
\Gamma_{\alpha}(\omega)\mathcal{G}^{\dagger}(\omega)\ket{\mathbf{r}}$, with
$\mathcal{G}(\omega)$ and $\Gamma_{\alpha}(\omega)$ the KS Green's function and
broadening matrices, respectively. Finally, the KS transmission function is 
$\mathcal{T}_{s, \alpha\alpha'}(\omega)=\text{Tr}\left\{\mathcal{G}_{s}(\omega)
\Gamma_{\alpha}(\omega)\mathcal{G}^{\dagger}(\omega)
\Gamma_{\alpha'}(\omega)\right\}$.

\subsection{Linear Response}
\label{linres}

Suppose we have a (multi-terminal) system in thermal eqilibrium characterized
by chemical potential $\mu$ and (common) temperature $T$ and we are interested
in the steady-state currents {\em to linear order} as external biases
and/or temperature gradients are applied to the system. In this linear regime,
the relationship between the currents $\textbf{I}$ and the external potentials
$\bm{\Phi}$ reads
\begin{gather}
  \textbf{I}= \textbf{L}\bm{\Phi}
  \label{eq_mult_relat_vary}
\end{gather}
with the $2({\cal{N}}-1)\times 2({\cal{N}}-1)$ conductance matrix $\textbf{L}$
and the current and potential vectors defined as
$\textbf{I}^{\intercal}=( I_{1},Q_{1},\dots,I_{\mathcal{N}-1},Q_{\mathcal{N}-1})$  and
$\bm{\Phi}^{\intercal}=(V_{1}/T, \Psi_{1}/T,\dots,V_{\mathcal{N}-1}/T,\Psi_{\mathcal{N}-1}/T)$,
respectively. By construction, the matrix elements of $\textbf{L}$ are defined
as
\begin{gather}
  \textbf{L}_{jk}= \frac{\partial \textbf{I}_j}{\partial \bm{\Phi}_k}
  \bigg\vert_{\bm{\Phi}=0}
  \label{eq_lmatrix_element}
\end{gather}
and from Onsager's relation it follows that $\textbf{L}$ is symmetric, i.e.,
$\textbf{L}_{j k}=\textbf{L}_{k j}$ with $k,j \in\{1,\ldots,2({\cal{N}}-1)\}$.

Since by construction, the KS currents equal the interacting ones to {\em any}
order, we may also linearize Eqs.~(\ref{eq_KS_densities_I}) and
(\ref{eq_KS_densities_Q}) to obtain
\begin{gather}
\textbf{I}=\textbf{L}_{s}\left(\bm{\Phi}+\bm{\Phi}_{\rm xc}\right) 
\label{eq_KS_changes}
\end{gather}
where $\textbf{L}_{s}$ is the non-interacting (KS) linear response matrix. 
To linear order, the changes in the xc potentials can be written as
\begin{gather}
  \bm{\Phi}_{\rm xc} = \textbf{F}_{\rm xc}\textbf{I}=
  \textbf{F}_{\rm xc}\textbf{L}\bm{\Phi}
  \label{eq_xc_changes}
\end{gather}
where we have defined the matrix of xc derivatives (which we alternatively
denote the xc kernel) $\mathbf{F}_{\rm xc}$ as 
\begin{gather}
	\mathbf{F}_{\rm xc}=\left.
	\left( {\begin{array}{cccccc}
	\frac{\delta V_{{\rm xc},1}}{\delta I_{1}} & \frac{\delta V_{{\rm xc},1}}{\delta Q_{1}}& \dots &\frac{\delta V_{{\rm xc},1}}{ \delta I_{\mathcal{N}-1}}  & \frac{\delta V_{{\rm xc},1}}{\delta Q_{\mathcal{N}-1}} \\
	\frac{\delta \Psi_{{\rm xc},1}}{\delta I_{1}} & \frac{\delta \Psi_{{\rm xc},1}}{\delta Q_{1}}& \dots &\frac{\delta \Psi_{{\rm xc},1}}{ \delta I_{\mathcal{N}-1}}  & \frac{\delta \Psi_{{\rm xc},1}}{\delta Q_{\mathcal{N}-1}}\\
	\vdots &\vdots& \ddots & \vdots&\vdots\\
\frac{\delta V_{{\rm xc},\mathcal{N}-1}}{\delta I_{1}} & \frac{\delta V_{{\rm xc},\mathcal{N}-1}}{\delta Q_{1}}& \dots &\frac{\delta V_{{\rm xc},\mathcal{N}-1}}{ \delta I_{\mathcal{N}-1}}  & \frac{\delta V_{{\rm xc},\mathcal{N}-1}}{\delta Q_{\mathcal{N}-1}} \\
\frac{\delta \Psi_{{\rm xc},\mathcal{N}-1}}{\delta I_{1}} & \frac{\delta \Psi_{{\rm xc},\mathcal{N}-1}}{\delta Q_{1}}& \dots &\frac{\delta \Psi_{{\rm xc},\mathcal{N}-1}}{ \delta I_{\mathcal{N}-1}}  & \frac{\delta \Psi_{{\rm xc},\mathcal{N}-1}}{\delta Q_{\mathcal{N}-1}}\\
	\end{array} } \right)
 \right |_{\substack{\textbf{I}=0}}
	\label{eq_matrix_F_xc}
\end{gather}
Combining Eqs.~(\ref{eq_mult_relat_vary}), (\ref{eq_KS_changes}), and
(\ref{eq_matrix_F_xc}), and using the fact that the $V_\alpha$ and $\Psi_\alpha$ are arbitrary,
we arrive at the Dyson equation for the many-body conductance matrix
$\mathbf{L}$ expressed in terms of the KS one $\mathbf{L}_{s}$
\begin{gather}
  \textbf{L}  = 
  \textbf{L}_{s}+\textbf{L}_{s}\textbf{F}_{\rm xc}\textbf{L},
  \label{eq_Dyson}
\end{gather}

Note that Eq.~(\ref{eq_Dyson}) expresses the many-body conductance matrix
completely in terms of iq-DFT quantities  
and may be rewritten as
\begin{align}
  \mathbf{F}_{\rm xc}=\mathbf{L}_{s}^{-1}-\mathbf{L}^{-1} \;.
  \label{eq_Dyson_XC}
\end{align}
As a consequence of $\mathbf{L}$ and $\mathbf{L}_s$ being symmetric, also
$\mathbf{F}_{\rm xc}$ must be symmetric,
$(\mathbf{F}_{\rm xc})_{k j}=(\mathbf{F}_{\rm xc})_{j k}$

\section{Single Impurity Anderson Model}
\label{siam}

In this section, we apply our multi-terminal formalism to the simplest quantum
thermal machine, namely, the Single Impurity Anderson Model (SIAM). This
system consists of a quantum  dot that can hold up to two interacting
electrons attached to $\mathcal{N}$ non-interacting electron reservoirs. The
Hamiltonian of the system reads

\begin{align}
  \hat{H} = &\sum_{\sigma}v \hat{n}_{\sigma} + U\hat{n}_{\uparrow}\hat{n}_{\downarrow}+ \sum_{\alpha k \sigma}\varepsilon_{\alpha k \sigma} \hat{c}^{\dagger}_{\alpha k \sigma} \hat{c}_{\alpha k \sigma}\nonumber\\
& + \sum_{k \alpha \sigma}\left( t_{\alpha k}\hat{c}^{\dagger}_{\alpha k \sigma}\hat{d}_{\sigma} + H.c.\right).
\label{eq_SIAM_H}
\end{align}

The first two terms in  Eq.~(\ref{eq_SIAM_H}) describe the isolated impurity,
with $v$ representing the on-site energy of the dot and $U$ symbolizing the
Coulomb interaction. The creation operators for electrons with spin $\sigma$
($\sigma=\uparrow,\downarrow$) in lead $\alpha$ and on the dot are denoted by
$\hat{c}_{\alpha k \sigma}^{\dagger}$ and $\hat{d}_{\sigma}^{\dagger}$, respectively.
The operators for the spin-resolved and the total density of electrons on the
dot are given by $\hat{n}_{\sigma}=\hat{d}^{\dagger}_{\sigma}\hat{d}_{\sigma}$ and
$\hat{n}=\hat{n}_{\uparrow}+\hat{n}_{\downarrow}$, respectively. The last term
of the Hamiltonian (\ref{eq_SIAM_H}) describes the tunneling between the dot
and the leads, with couplings $\Gamma_{\alpha}(\omega) =
2 \pi \sum_k |t_{\alpha k}|^2 \delta(\omega -\varepsilon_{k \alpha})$. We work in
the wide band limit (WBL), i.e., the leads are assumed to be featureless and
described by frequency-independent couplings
$\Gamma_{\alpha}(\omega)=\gamma_{\alpha}$ (with
$\alpha \in\{1,\ldots,\mathcal{N} \}$). Without loss of generality, from here
onwards we set the chemical potential $\mu=0$. 

\subsection{Many body model for the construction of the exchange
  correlation kernel}

In order to apply our multi-terminal iq-DFT framework, approximations for the
xc potentials need to be constructed. Here we restrict ourselves to the
linear-response regime and therefore we only need to construct the xc kernel
matrix $\mathbf{F}_{\rm xc}$.

As in previous works \cite{stefanucci2015steady,sobrino2021thermoelectric}, a
useful starting point for this construction is to write the interacting
density on the dot as well as the (particle and heat) currents in the leads
in terms of the many-body spectral function $A(\omega)$ as
\begin{subequations}
  \begin{align}
    &n=2 \sum_{\alpha=1}^{\cal{N}}\int \frac{\gamma_{\alpha}}{\gamma}
    f(\omega_{\alpha})A(\omega) \label{n_MB_SIAM}\\
    &I_{\alpha}=2
    \sum_{\alpha'=1}^{\cal{N}}\frac{\gamma_{\alpha}\gamma_{\alpha'}}{\gamma}
    \int \left[f(\omega_{\alpha})-f(\omega_{\alpha'})\right]A(\omega)\\
    &Q_{\alpha}=2
    \sum_{\alpha'=1}^{\cal{N}}\frac{\gamma_{\alpha}\gamma_{\alpha'}}{\gamma}
    \int \left[f(\omega_{\alpha})-f(\omega_{\alpha'})\right](\omega-V_{\alpha})
    A(\omega)
  \end{align}
  \label{eqs_MB_SIAM}
\end{subequations}
with $\gamma = \sum_{i=1}^{\cal{N}} \gamma_{\alpha}$ and $\omega_{\alpha} =
\frac{\omega - V_{\alpha}}{1+ \Psi_{\alpha}}$.

In order to proceed, we consider the following model for the many-body
spectral function (MBM) which can be derived from the equations of motion
technique \cite{haug2008quantum} and provides a reasonably accurate
approximation for $T/\gamma >1$, i.e., for the Coulomb blockade regime
\cite{kurth2017transport} 
\begin{align}
  A(\omega)=\frac{\gamma\left(1-\frac{n}{2}\right)}{(\omega-v)^{2} +
    \frac{\gamma^2}{4}}
  +\frac{\gamma\frac{n}{2}}{(\omega-v-U)^{2}+\frac{\gamma^2}{4}}.
  \label{eq_model_specfunc_SIAM}
\end{align}
For the non-interacting case with gate potential $v_s$, this spectral
function becomes
\begin{equation}
  A_s(\omega)= \frac{\gamma}{(\omega-v_s)^{2}+\frac{\gamma^2}{4}} \;.
  \label{eq_KS_model_specfunc}
\end{equation}
Then all the integrals for density and currents can be evaluated
analytically \cite{sobrino2021thermoelectric} with the results
\begin{subequations}
  \begin{align}
    n^s&=1 - \frac{2}{\pi\gamma}\sum_{\alpha=1}^{\cal{N}} \gamma_{\alpha}
    \text{Im}\left[\psi(z_{z,\alpha})\right]\label{eq_n_SIAM}\\
    I^s_{\alpha}&=\frac{2\gamma_{\alpha}}{\pi\gamma}\left[
      \sum_{\alpha'=1}^{\cal{N}}\gamma_{\alpha'}\left(\text{Im}
      \left[\psi(z_{s,\alpha'})\right]-
      \text{Im}\left[\psi(z_{s,\alpha})\right]\right)\right]\label{eq_I_SIAM}\\
    Q^s_{\alpha}&=\frac{\gamma_{\alpha}}{\pi}\left[
      \sum_{\alpha'=1}^{\cal{N}}\gamma_{\alpha'}\left(\text{Re}
      \left[\psi(z_{s,\alpha})\right]-
      \text{Re}\left[\psi(z_{s,\alpha'})\right]\right)\right]\nonumber\\
    & + \frac{ \gamma_\alpha}{\pi}\sum_{\alpha'=1}^{\cal{N}}\left[
      \gamma_{\alpha'} \log\left(\frac{1+\Psi_{s,\alpha}}{1+\Psi_{s,\alpha'}}\right) \right]
    +(v_{s}-V_{s,\alpha})I_{\alpha}
     \label{eq_Q_SIAM}
  \end{align}
  \label{eqs_densities_SIAM}
\end{subequations}
where $z_{s,\alpha}=\frac{1}{2}+\frac{\frac{\gamma}{2}+i(v_{s}-V_{s,\alpha})}
{2\pi T (1+ \Psi_{s,\alpha})}$ and $\psi(z)$ is the digamma function with
general complex argument $z$.\cite{abramowitz1965handbook}
Eqs.~(\ref{eq_I_SIAM}) and (\ref{eq_Q_SIAM}) can be expanded to linear order
in the biases $V_{s,\alpha}$ and temperature gradients $\Psi_{s,\alpha}$ and the
resulting integrals for the expansion coefficients can also be evaluated
analytically. Writing $j= 2\alpha-1$ ($j=2 \alpha$) for $j$ odd (even) and
similarly $k= 2\alpha'-1$ ($k=2 \alpha'$) for $k$ odd (even) we can express
the non-interacting conductance matrix $\mathbf{L}_s(v_s)$ as function
of $v_s$ in the compact form
\begin{equation}
  \mathbf{L}_{s, jk}(v_s) = \tilde{\gamma}_{jk} M_{jk}(v_s)
  \label{lmatrix_nonint}
\end{equation}
where the prefactor $\tilde{\gamma}_{jk}$ only depends on the couplings
to the leads and can be written as
\begin{equation}
  \tilde{\gamma}_{jk} = 2 \gamma_{\alpha} \left(
  \delta_{\alpha,\alpha'} + \frac{\gamma_{\mathcal{N}} - \gamma_{\alpha'}}{\gamma} \right)
\end{equation}
On the other hand, the coefficient $M_{jk}(v_s)$ depends on $v_s$ and
is defined as
\begin{equation}
  M_{jk}(v_s) = \left\{
  \begin{array}{cl}
    J_0(v_s) & \mbox{for both $j,k$ odd} \\
    J_2(v_s) & \mbox{for both $j,k$ even} \\
    J_1(v_s) & \mbox{otherwise}
  \end{array}
  \right.
\end{equation}
where
\begin{equation}
  J_l(v_s) = - \int \omega^l f'(\omega)
  \frac{\gamma}{(\omega-v_s)^2 + \frac{\gamma^2}{4}}
  \label{def_jk}
\end{equation}
where $f'(\omega)= \frac{{\rm d}}{{\rm d} \omega} f(\omega)$. The integrals
of Eq.~(\ref{def_jk}) can also be computed analytically
\cite{sobrino2021thermoelectric} with the results
\begin{subequations}
  \begin{align}
    J_0(v_s) & = \frac{1}{2 \pi^2 T} {\rm Im}(i \psi^{(1)}(z_1^s)) \\
    J_1(v_s) & = \frac{1}{2 \pi^2 T} {\rm Im}(z_0^s \psi^{(1)}(z_1^s)) \\
    J_2(v_s) & = - \frac{\gamma}{4 \pi^2 T} {\rm Re}(z_0^s \psi^{(1)}(z_1^s))
    + v_s J_1(v_s) + \frac{\gamma}{2 \pi}
  \end{align}
\end{subequations}
where $z_0^s= \frac{\gamma}{2} + i v_s$, 
$z_1^s = \frac{1}{2} + \frac{z_0^s}{2 \pi T}$ and $\psi^{(1)}(z)$ is the
trigamma function.\cite{abramowitz1965handbook}

Due to the simple structure of the model many-body spectral function
of Eq.~(\ref{eq_model_specfunc_SIAM}), we can now express the corresponding
{\em interacting} conductance matrix in terms of quantities obtained for
the non-interacting case as
\begin{equation}
  \mathbf{L}_{jk}(v) = \tilde{\gamma}_{jk} \left[ \left( 1 - \frac{n}{2} \right)
    M_{jk}(v) + \frac{n}{2} M_{jk}(v+U) \right] \;.
  \label{lmatrix_int}
\end{equation}
Eqs.~(\ref{lmatrix_int}) and (\ref{lmatrix_nonint}) provide the analytical
forms of the conductance matrices in terms of the gate potential $v_s$ and
$v$, respectively. In order to express the xc kernel
matrix $\mathbf{F}_{\rm xc}$ as functional of the density, we still need to
express these gate potentials in terms of the density.
 The interacting (non-interacting) gate-density
relation $v(n)$ ($v_s(n)$) is obtained by numerically inverting the density-potential
relationship obtained by inserting the spectral function
Eq.~(\ref{eq_model_specfunc_SIAM}) (Eq.~(\ref{eq_KS_model_specfunc})) into Eq.~(\ref{n_MB_SIAM}). 
From Eq.~(\ref{eq_Dyson_XC}), the xc kernel matrix can then be expressed
solely in terms of the density as
\begin{equation}
\mathbf{F}_{\rm xc}(n) = \mathbf{L}_s(v_s(n))^{-1} - \mathbf{L}(v(n))^{-1} \;.
\end{equation}

In order to complete our DFT scheme we need to provide an approximation for
the Hxc (gate) potential $v_{\rm Hxc}(n)$ to be used in \cref{eq_Dyson_XC}.
In the present work we use the Hxc potential of the single site
model\cite{stefanucci2011towards} (SSM) with an effective temperature $T^{*}$ as
proposed in Ref.~\onlinecite{sobrino2019steady} in order to correctly account
for the dependence on the coupling.  We will refer to this scheme as $T^*$ DFT. In particular, we replace the effective
temperature of Ref.~\onlinecite{sobrino2019steady} with 
\begin{subequations}
\begin{gather}
	T^{*}(T,\gamma, U)=\frac{T^{2}+\eta(\gamma,U,T)\gamma T+\eta^2(\gamma,U,T)\gamma^2}{T+\eta(\gamma,U,T)\gamma},\\
	\eta(\gamma,U,T)=\eta_1(\gamma/T)\eta_2(U/T),\\
	\eta_1(x)=0.478x^{-\frac{1}{2}}+0.1331,\label{eq_eta1}\\
	\eta_2(x)=0.676\arctan{(0.064x)}+0.661.\label{eq_eta2}
\end{gather}
\label{eq_T_star}
\end{subequations}

 To assess the accuracy of the proposed analytical Hxc functional, we compare it against the numerically exact inversion or reverse-engineered (RE) Hxc functional. This is computed as the difference between the numerically inverted interacting and non-interacting gate potentials from \cref{eqs_densities_SIAM,eqs_MB_SIAM} at equilibrium.  We have found that the above expressions, with a fit for the numerical values in \cref{eq_eta1,eq_eta2}, correct the standard DFT results and bring them closer to the MBM for the SIAM.

\begin{figure}
  \includegraphics[width=0.49\linewidth]{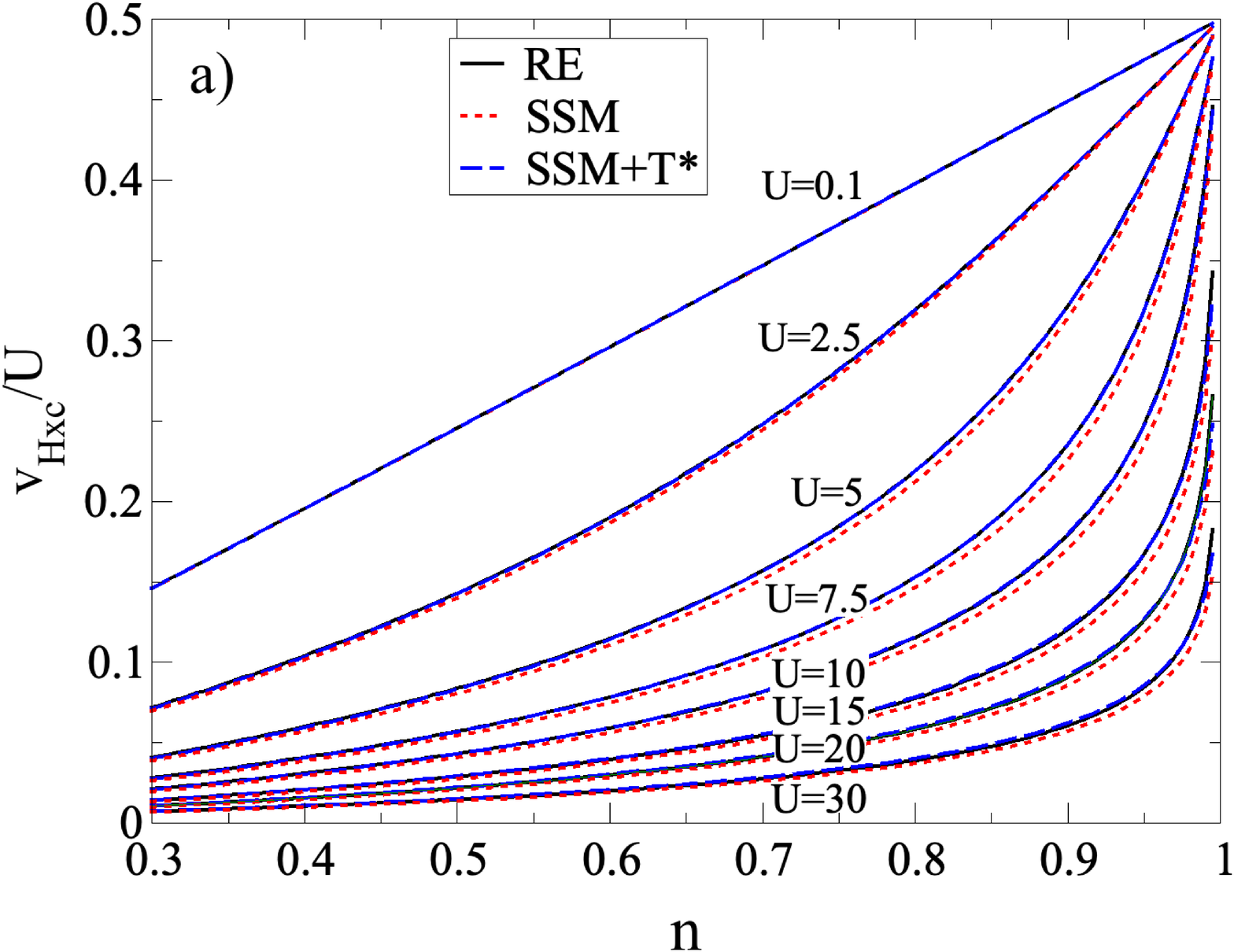}
  \includegraphics[width=0.49\linewidth]{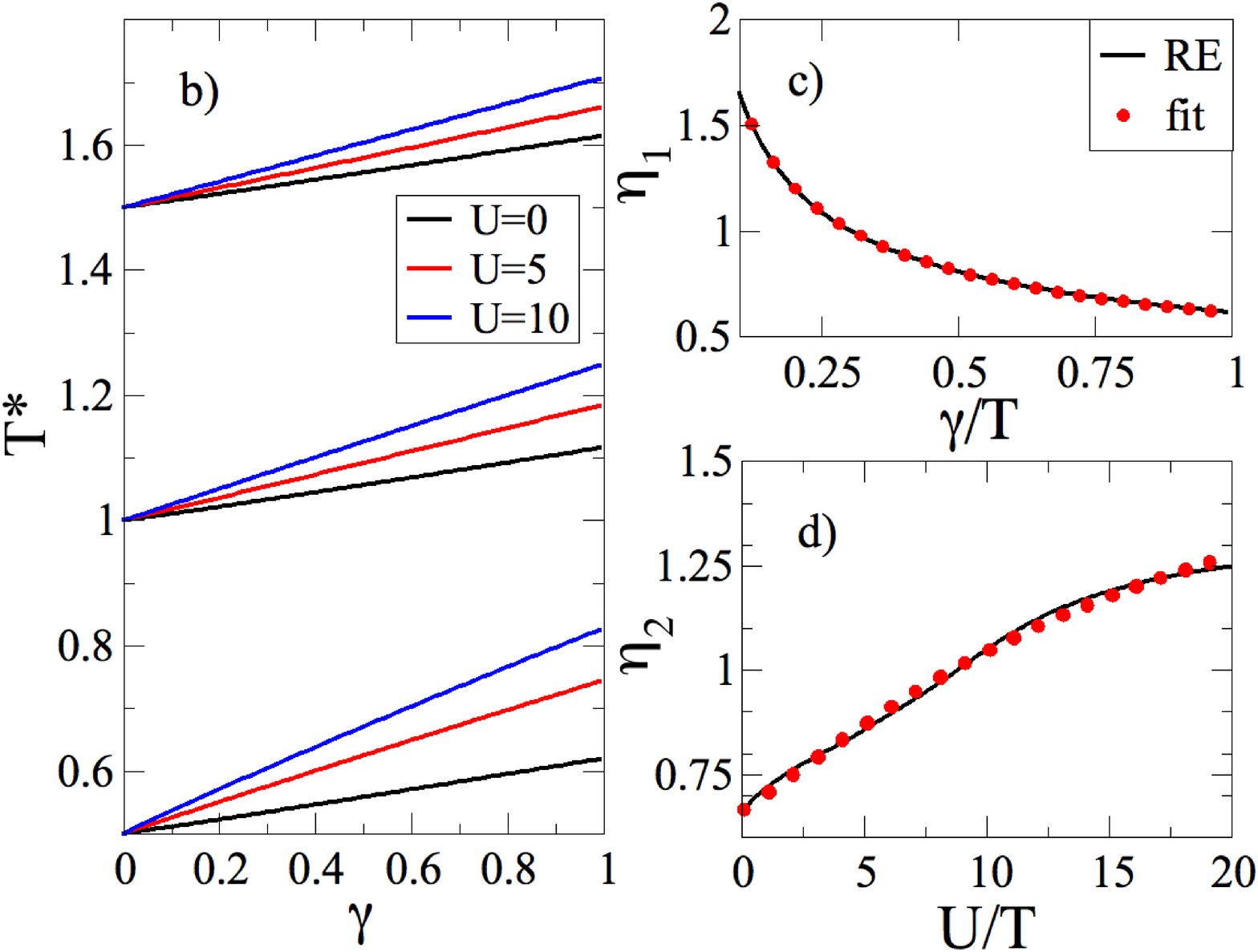}
  \caption{a) Comparison of the reverse engineered (RE) Hxc potential from the
    many-body model with the one of the single-site model (SSM) and the SSM
    corrected with the effective temperature \cref{eq_T_star} for different
    Coulomb interactions and $\gamma/T=0.1$. b) Effective temperature $T^*$
    for $T=0.5,1,1,5$ as function of the coupling $\gamma$.
    c) and d) Comparison between the rate equation fitting functions $\eta_1$ and
    $\eta_2$ and the parametrizations of \cref{eq_eta1,eq_eta2}.
    }
  \label{fig_2}
\end{figure}

In Fig.~\ref{fig_2} a) we compare the Hxc potential $v_{Hxc}$ from
reverse-engineerings of our many-body model with the SSM Hxc potential (no
coupling) and the SSM Hxc potential with effective temperature to take into
account the coupling for different values of the Coulomb interaction. The
correction induced by the effective temperature for different couplings
(\cref{fig_2} b)) produces a small variation in the Hxc potential which is
essential to capture the correct strong-interaction limit of the thermoelectric
efficiencies which will be presented in the next section. This dependence on
subtle details of the Hxc functional is reminiscent of the ones observed in
Ref.~\onlinecite{sobrino2022level} for the description of the level occupation
switching effect.

\subsection{Results}
\label{results_num}

In this section we present our numerical results. Here our interest lies in the
description of the thermoelectric efficiency $\eta$ as well as various
linear-response transport coefficients
of the quantum thermal machine for finite Coulomb interactions and multiple
reservoirs, see \cref{fig_1} for $\mathcal{N}=3$. 

In order to access these quantities, we first solve the DFT problem in the
standard way to obatain the density. Then, following the scheme presented in
the previous section we compute the kernel matrix $\mathbf{F}_{\rm xc}$, the
linear response matrix $\mathbf{L}$ and finally, through
\cref{eq_mult_relat_vary}, the currents to linear order. The
multi-terminal efficiency can then be obtained from the currents
through \cite{erdman2017thermoelectric,mazza2014thermoelectric}
\begin{align}
  \eta = \frac{P}{\sum_{\alpha_+}Q_{\alpha}}=\frac{\sum_{\alpha=1}^{\mathcal{N}}Q_\alpha}{\sum_{\alpha_+}Q_{\alpha}}
  \label{eq_efficiency}
\end{align}
where the symbol $\sum_{\alpha_+}$ indicates that the sum is restricted to
positive contributions of the heat currents. \cref{eq_efficiency} is
restricted to positive values of the output power
$P$.  In the following, we assume the condition $P>0$ is always satisfied.
Usually, this efficiency is presented
normalized with its upper bound, the Carnot efficiency. The Carnot efficiency
in turn is obtained by imposing zero entropy production
$\dot{\mathcal{S}} = \frac{1}{T}\sum_{\alpha=1}^{\mathcal{N}-1}\left( I_{\alpha}
V_{\alpha} + Q_{\alpha}  \Psi_{\alpha} \right) =0$ which leads to
\begin{align}
  \eta_C =\frac{\sum_{\alpha=1}^{\mathcal{N}-1}Q_\alpha(1-\frac{T_\mathcal{N}}{T_\alpha})}
      {\sum_{\alpha_+}Q_{\alpha}}
      \label{eq_Carnot_efficiency}.
\end{align}
For the two terminal case one recovers the well known expression
$\eta_C=1-T_2/T_1$.\cite{mitchison2019quantum} 

\begin{figure}
  \centering
  \includegraphics[width=\linewidth]{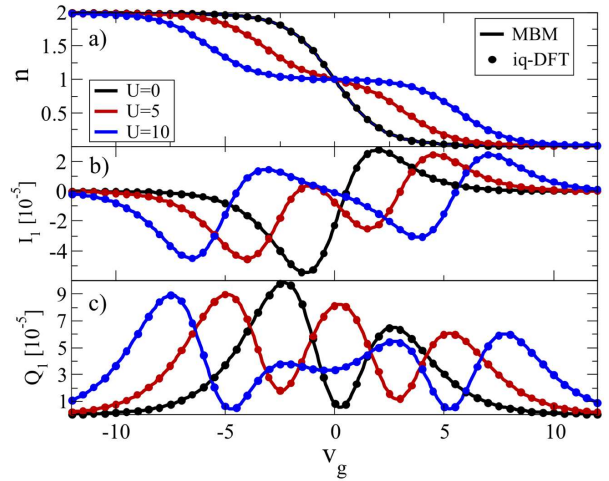}
  \caption{Comparison of the density and currents $I_1$ and $Q_1$ from
    the reference MBM and iq-DFT as function of the gate voltage
    $v_g=v+\frac{U}{2}$ for different Coulomb interactions and $\mathcal{N}=3$. The parameters used are $\gamma_i=\gamma/3=0.1T$, $V_1=-V_2=-5\cdot10^{-4}T$, $\Psi_1=-\Psi_2=10^{-3}$ and $V_3=\Psi_3=0$.}
  \label{fig_3}
\end{figure}

In the following, unless explicitly noted, all energies are given in units of the temperature $T$. In \cref{fig_3} we present a comparison of the density and currents from the
many-body model and iq-DFT. The small values used for the potentials
$V_1=-V_2=-5\cdot10^{-4}T$, $\Psi_1=10^{-3}$ and $V_3=\Psi_2=\Psi_3=0$ ensure
the applicability of the linear response equations for the currents
\cref{eq_mult_relat_vary}. The agreement between MBM and iq-DFT is excellent
both for the non-interacting as well as for the (strongly) correlated case,  although the agreement may decrease away from the CB regime.

\begin{figure}
  \centering
  \includegraphics[width=\linewidth]{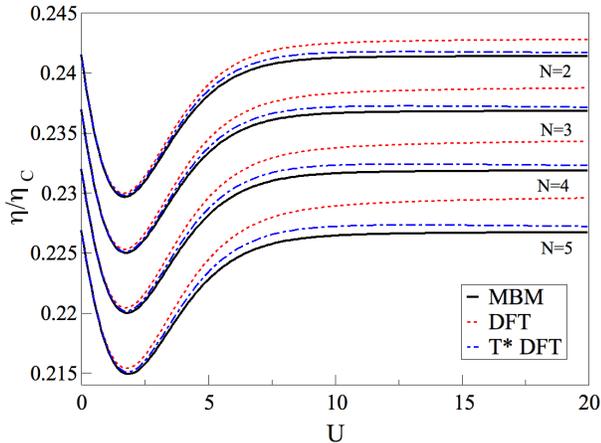}
  \caption{Thermoelectric efficiency normalized by the Carnot efficiency
    as function of the Coulomb interaction for different number of leads
    $\mathcal{N}=2,3,4,5$. Comparison of the results of the many-body model (MBM) with
    iq-DFT results obtained with the original SSM Hxc potential of
    Ref. \onlinecite{stefanucci2011towards} (DFT) and those with the modified SSM using
    the effective temperature $T^*$ of \cref{eq_T_star} ($T^*$ DFT). The parameters are $\gamma_i=\gamma/3=0.1T$, $v=2$, $V_1=-5\cdot10^{-4}T$, $V_i=-V_1/(\mathcal{N}-2)$ for
    $i=2,\ldots,\mathcal{N}-1$,  $\Psi_1=10^{-3}$,
    $\Psi_i=-\Psi_1/(\mathcal{N}-2)$ for $i=2,\ldots,\mathcal{N}-1$ and
    $\Psi_{\mathcal{N}}=V_{\mathcal{N}}=0$.
  }
  \label{fig_4}
\end{figure}

In \cref{fig_4} the thermoelectric efficiency is shown as function of the
Coulomb interaction for $\gamma_i=\gamma/3=0.1T$, $v=2T$. We observe that in
the limit of very strong interaction the efficiency exactly corresponds to the
non-interacting one. This can be understood from our many-body model, i.e.,
inserting the model spectral function of Eq.~(\ref{eq_model_specfunc_SIAM})
into Eqs.~(\ref{eqs_MB_SIAM}). In the strong-interaction limit the
contribution of the pole of the spectral function at $v+U$  is negligible for
all ``densities'' due to the integral cutoff of the Fermi functions.
Therefore, the density and currents in this limit can be rewritten as 
\begin{subequations}
	\begin{gather}
		n_{U\to\infty}=\frac{n^s(v,\mathbf{\Phi})}{1+\frac{1}{2}n^s(v,\mathbf{\Phi})}\label{eq_n_MBM}\\
		I_{U\to\infty}=I^s(v,\mathbf{\Phi})(1-\frac{1}{2}n^s(v,\mathbf{\Phi}))\\
		Q_{U\to\infty}=Q^s(v,\mathbf{\Phi})(1-\frac{1}{2}n^s(v,\mathbf{\Phi})),
	\end{gather}
	\label{eqs_MBM_currents_U_to_infinity}
\end{subequations}
where $n^s(v,\Phi)$, $I^s(v,\Phi)$ and $I^s(v,\Phi)$ are the non-interacting
expressions of Eq.~(\ref{eqs_densities_SIAM}) evaluated at gate $v$ and
potentials $\Phi$. \cref{eqs_MBM_currents_U_to_infinity} explicitly shows that in the strong-interacting limit, the interacting density and currents are fully determined by their non-interacting versions evaluated at the interacting potentials. Inserting \cref{eqs_MBM_currents_U_to_infinity} into
\cref{eq_efficiency} one finds that the prefactor $(1-\frac{1}{2}n^s(v,\mathbf{\Phi}))$
cancels out and we recover the efficiency of the non-interacting limit.  For
the parameters studied, in \cref{fig_4} the efficiency decreases
as the Coulomb interaction is increased, finding the minimum around $U\sim2$.
Then it increases again up to the non-interacting value. The correction of the
effective temperature of \cref{eq_T_star} is relevant in the strong-interacting
limit: while with the original SSM parametrization of the Hxc potential, in the strongly correlated limit the efficiency does not approach the MBM limit, with the new parametrization it does.
 Our $T^*$-DFT results agree well with the MBM results in the region of parameters we investigated. We point out that the small deviation between $T^*$-DFT and MBM, in \cref{fig_4} stems from similar discrepancies in the heat and electrical currents, which are difficult to detect in Fig. 3.

\begin{figure}
  \centering
  \includegraphics[width=\linewidth]{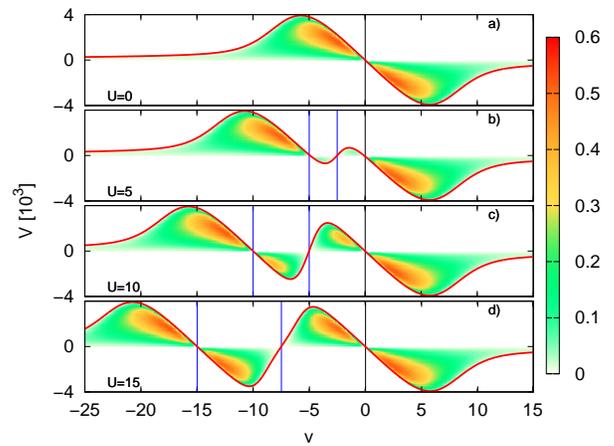}
  \caption{Thermoelectric efficiency normalized by Carnot efficiency as
    function of the gate level and the bias potential $V$ for the
    configuration $V_1=-V_2=V$, $V_{3}=0$  $\Psi_1=-\Psi_2=10^{-3}$,
    $\Psi_{3}=0$ and parameters $\gamma_1=\gamma_2=\gamma_3=0.1T$. From a) to
    d) the Coulomb interaction is $U=0,5,10,15$. The red line corresponds to
    the open circuit voltage $V_{oc}$ and the blue vertical lines correspond
    to the gate values $v=-U/2,-U$ at which $V_{oc}=0$ .}
  \label{fig_5}
\end{figure}

In \cref{fig_5}, the efficiency is calculated for
$\mathcal{N}=3$ as function of the gate
$v$ and bias $V$ for the configuration  $V_1=-V_2=V$, $V_{3}=0$,
$\Psi_1=-\Psi_2=10^{-3}$, $\Psi_{3}=0$ and different values of the Coulomb
interaction $U/T=0,5,10,15$ from a) to d), respectively. The red line
represents the open-circuit voltage $V_{oc}$, and corresponds to the bias at
which the ouput power is zero, which in our configuration corresponds to the bias at which $I_1=I_2$. 

In \cref{fig_5}a) the (iq-DFT) efficiency is presented for the three terminal
setup in the non-interacting case. As in the two terminal
case\cite{nakpathomkun2010thermoelectric}, the efficiency acts as a power
generator for voltages $V_{oc}<V<0$ (region 1) if the gate is negative and for
voltages $0<V<V_{oc}$ (region 2) if the applied gate is positive, and the open
circuit voltage only vanishes at $v=0$. The application of a finite Coulomb
interaction in the QD (\cref{fig_5}b),c) and d)) produces two new regions
(where the QD acts as a power generator) that emerge in between regions 1
and 2. While the shape of regions 1 and 2 remain unchanged, region 1 is
shifted to $v-U$. The new regions at finite U appear at
$V_{oc}<V<0$, $-U<v<-U/2$ (region 3) and $0<V<V_{oc}$, $-U/2<v<0$ (region 4).
The gate values at which there is a new transition between the regions is
represented by the blue lines in the plots and correspond to the new gates at
which the open circuit voltage vanishes $v=-U/2,-U$.
As the Coulomb interaction is increased, the area of regions 3 and 4 increases
and so the value of the efficiency inside these regions. In the high
interaction limit the regions 1 and 3 and the regions 2 and 4 become
equivalent, recovering the non-interacting limit around the Fermi energy. The
same behavior has been observed when the number of leads is $\mathcal{N}>3$
(not shown).
 
\begin{figure}
  \centering
  \includegraphics[width=\linewidth]{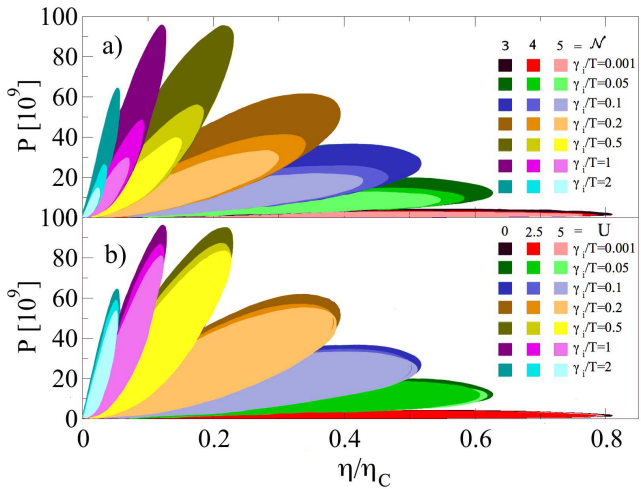}
  \caption{Output power, in atomic units, as function of the efficiency for different total
    coupling strength $\gamma$ and a) different number of reservoirs for the
    non-interacting QD,  $U=0$ and b) different Coulomb interactions $U$ for the three terminal case $\mathcal{N}=3$. Each solid region is obtained by scanning all
    the possible gate and bias combinations where the QD acts as a
    thermoelectric generator. The bias configuration selected corresponds
    to $V_1=-(\mathcal{N}-1)V_i$ with $i=1,\ldots,\mathcal{N}-1$ and
    $V_{\mathcal{N}}=0$.}
  \label{fig_6}
\end{figure}

We now focus our attention on the influence of the total coupling $\gamma$ on
the output power $P$ and the efficiency $\eta$.
The application of a bias inside the regions previously
defined  for different gates, results in a loop of the output power as
function of the bias. \cite{josefsson2018quantum}
Scanning of all possible gates fills up the regions shown in \cref{fig_6}.
We observe that for the different numbers of leads studied here (\cref{fig_6} a)), the
efficiency always approaches the Carnot efficiency as the coupling strength
is reduced since the transmission function in this case approaches a delta
function.\cite{humphrey2002reversible,josefsson2018quantum} On the other hand,
when the coupling strength is increased, the number of electrons which
contribute to the power generation increase, and, therefore, the output power
reaches its maximum around $\gamma_{max}\sim0.5T$.  For couplings larger than
$\gamma_{max}$, the transmission function $\mathcal{T}_{\alpha\alpha'}(\omega)=
\gamma_{\alpha}\gamma_{\alpha'}/\gamma A(\omega)$ in \cref{eqs_densities_SIAM}
allows more energy states to contribute, in particular some  negative
contributions of the difference
$\tilde{f}_{\alpha}(\omega)-\tilde{f}_{\alpha'}(\omega)$ which decrease the
output power. It is worth noting that the larger the coupling to the leads, the larger the ratio between the $\mathcal{N}$ and $\mathcal{N}+1$ regions areas becomes. In  \cref{fig_6} b) the thermoelectric efficiency is presented for different values of the Coulomb interaction and several coupling strengths. For small values of the coupling to the leads, the efficiency decreases as the interaction is applied and then it increases again tending to the non-interacting value, while the output power remains essentially unchanged. For larger values of the coupling, the efficiency remains mostly unchanged while the output power starts to decrease with the interaction.

\begin{figure}
  \centering
  \includegraphics[width=\linewidth]{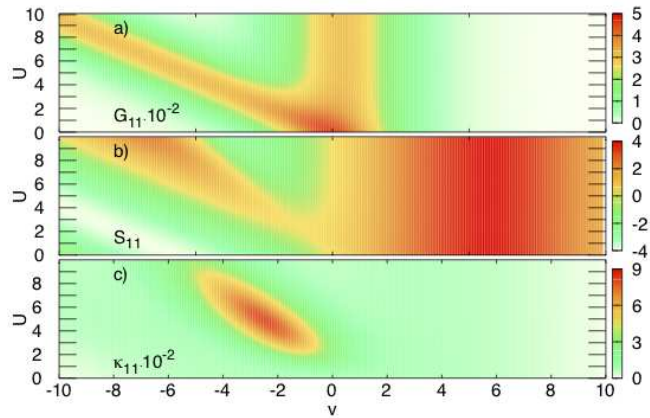}
  \caption{ Transport coefficients $G_{11}$, $S_{11}$, and $\kappa_{11}$
     as function of the Coulomb interaction and
    the gate level for $\mathcal{N}=3$. The coupling strengths considered are
    $\gamma=3\gamma_i=0.1T$. All quantities expressed in atomic units.}
  \label{fig_7}
\end{figure}

To conclude our multi-terminal study of the quantum machine, we apply our
formalism to calculate the transport coefficients and the figure of merit for
the case  $\mathcal{N}=3$ and equal couplings $\gamma_i$. In \cref{fig_7} we show the transport
coefficients $G_{11}$, $S_{11}$, $\kappa_{11}$  as function of
the Coulomb interaction and the gate level following the definitions of these coefficients derived in Ref.~\onlinecite{mazza2014thermoelectric}.
 Note that from these definitions, in the symmetric coupling setup $\gamma_1=\gamma_2=\gamma_3$ one can analytically show that $G_{11}=G_{22}$, $S_{11}=S_{22}$, $\kappa_{11}=\kappa_{22}$ while all the other off-diagonal transport coefficients vanish. 

The electrical conductance (\cref{fig_7} a)) $G_{11}=G_{22}$ is
maximum  in the non-interacting case for the gate $v=0$. 
At finite U, as expected, this feature splits in two Coulomb blockade
peaks at $v=0$ and at $v=-U$, a direct consequence of the form of the MBM
spectral function of \cref{eq_model_specfunc_SIAM}. 

The Seebeck coefficient $S_{11}$ (\cref{fig_7} b)) presents two main features
which evolve with the Coulomb interaction: at negative gates $-10-U<v<-U$ the
Seebeck coefficient has its minimum and at positive gates $0<v<10$ the Seebeck
coefficient
evolves to its maximum value. For strong correlations $U/T\gtrsim 5$, a new
feature appears between the other two structures alternating positive and
negative contributions.
Finally, the  thermal conductance  $\kappa_{11}$  shows a localized structure
distributed along gates  $v\approx-U/2$ and reaching its maximum around
$U\sim 5$, see \cref{fig_7} c).

\begin{figure}
  \centering
  \includegraphics[width=\linewidth]{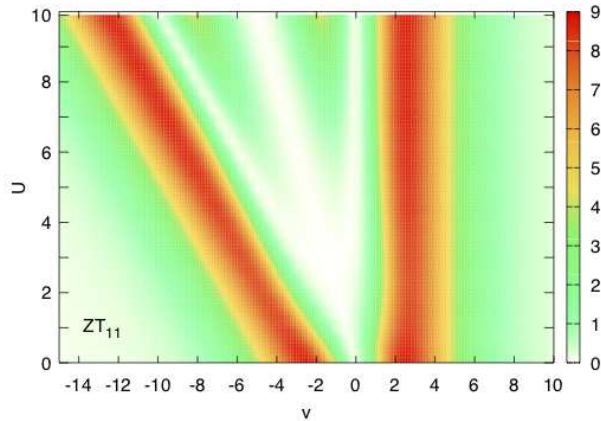}
  \caption{Diagonal element of the figure of merit as function of the
    Coulomb interaction and the gate level for $\mathcal{N}=3$. The
    coupling strengths considered are  $\gamma=3\gamma_i=0.1T$}
  \label{fig_8}
\end{figure}

The figure of merit in the multi-terminal setup can be evaluated from
\begin{align}
ZT_{ij}= \frac{TG_{ij}^{2}S_{ij}}{\kappa_{ij}}.
\end{align}
In  \cref{fig_8} the diagonal element of the figure of merit is presented as
function of Coulomb interaction and gate level. $ZT_{11}=ZT_{22}$ is mostly dominated by $S_{11}=S_{22}$. It shows two stripe regions
centered at $v=-2.5-U,2.5$ where the figure of merit is maximized. 

\section{Conclusions}
\label{conclus}

In this work we generalize the iq-DFT theory for the description of electronic
and thermal transport through nanoscale junctions connected to an arbitrary
number of electrodes. The theory is established under a one-to-one
correspondence between the set of ``densities''
($n, I_{1},Q_{1},\dots,I_{\mathcal{N}-1}, Q_{\mathcal{N}-1}$)  and the set of
``potentials'' ($v, V_{1}/T,\Psi_{1}/T,\dots,V_{\mathcal{N}-1}/T,\Psi_{\mathcal{N}-1}/T$) in a finite domain around the equilibrium state. The KS system requires $2 (\mathcal{N}-1)$ xc potentials which need to be parametrized.

We derived the linear response of the (multi-terminal) formalism finding
formally exact expressions for the linear response electrical and heat
currents, the figure of merit, the thermoelectric efficiency and the many-body
transport coefficients, i.e., the electrical conductances, the Seebeck
coefficients, as well as the thermal conductances. These quantities are fully
and exactly expressed purely in terms of quantities accessible to the
iq-DFT framework, i.e., the xc kernel matrix $\mathbf{F}_{xc}$ and the Hxc
potential $v_{\rm{Hxc}}(n)$.

We applied the framework to an interacting quantum thermal machine with three,
four and five reservoirs in the linear response and the Coulomb Blockade
regime. We constructed the xc kernel matrix from reverse engineering of a
many-body model, finding excellent agreement with the the reference many-body
results for the currents and the transport coefficients as well as
thermoelectric efficiency. We have found and identify the regions where the
system acts as a thermal generator for different Coulomb interactions and
analyzed these regions against the output power for several couplings to the
leads and different number of reservoirs. 
Moreover, we understood analytically that in the strong-interaction limit
the thermoelectric efficiency exactly corresponds to the non-interacting one.

\section{ACKNOWLEDGMENTS}
We acknowledge financial support through Grant PID2020-112811GB-I00 funded by
MCIN/ AEI/10.13039/501100011033  as well as by grant IT1453-22 “Grupos
Consolidados UPV/EHU  del Gobierno  Vasco”.

\end{document}